\documentstyle[prb,aps,twocolumn]{revtex}
\begin{document}
\renewcommand{\bottomfraction}{1.0}
\renewcommand{\topfraction}{1.0}
\renewcommand{\textfraction}{0.0}
\renewcommand{\floatpagefraction}{0.9}
\narrowtext

\noindent{\bf Phase coherence and ``fragmented'' Bose condensates}
 
\noindent{D.S. Rokhsar}

\noindent{\it Department of Physics, University of California,
Berkeley,  CA 94720-7300, USA and Physical Biosciences Division,
Lawrence Berkeley National Laboratory, Berkeley, CA 94720, USA}

\medskip
 
\noindent{submitted November 27, 1998; PACS 03.75 Fi}

\bigskip

\noindent{\bf We show that a ``fragmented Bose condensate'' in which
two  or more distinct single-particle states are macroscopically
occupied by the  same species of boson is inherently unstable to the
formation of a conventional Bose condensate whose macroscopically
occupied state is a linear combination of the  ``fragments'' with
definite relative phases.  A related analysis shows that a reproducible
relative phase develops when two initially decoupled  condensates are
placed in contact.}

\bigskip

The essential property of a Bose condensate is the existence of a single
macroscopically large eigenvalue of its one-body density
matrix.\cite{Penrose,Beliaev,Yang}  This property reveals the existence
of long-range phase coherence across the fluid, which leads to the
two-fluid description of Bose condensates and implies the quantization
of circulation in superfluid flow.\cite{Anderson,Anderson2} 
Hypothetical Bose states  without this property were discussed long ago
by Nozieres and St. James \cite{Nozieres-StJames,Nozieres}, and have
recently received renewed theoretical attention in the case of trapped
bosons.\cite{Wilkin,Pethick}  In particular,  Wilkin {\em et
al.}\cite{Wilkin} found that the azimuthally symmetric rotating states
of a dilute Bose gas with attractive interactions are ``fragmented''
Bose condensates in which many single-particle states (of differing
angular momenta) are macroscopically occupied.  Such multi-condensate
Bose fluids would be expected to have unusual properties.

A related set of issues stems from a {\em gedankenexperiment} proposed
by Anderson,\cite{Lesson}  in which two previously isolated Bose
condensates are brought into particle-exchanging contact.  Before
contact is initiated, each isolated subsystem has its own
macroscopically occupied state. The two isolated Bose subsystems
considered together therefore provide a simple example of a system with
a ``fragmented'' condensate.  What happens when the condensates are
allowed to exchange particles -- will the phase difference be random
from experiment to experiment,\cite{Lesson,JY,CD} or be reproducibly
zero\cite{LeggettSols,Leggett}?

Here we show that when fragmented condensates overlap, they are
inherently unstable to the formation of a single, conventional
condensate of well-defined relative phase.  The essential idea is that
even weak Josephson coupling between the fragments rapidly generates
phase coherence, modifying the density matrix deterministically to give
a unique macroscopically occupied single particle state.   The proof
proceeds by first assuming the existence of a Bose state with multiple
condensates, and then explicitly constructing a lower energy state. 
The argument is very general, and applies to strongly as well as weakly
interacting systems, and for either sign of the scattering
length.  As a corollary, we show that the relative phase that emerges
when two previously uncoupled condensates are placed in contact is
reproducible, and is determined by the physics of coherent particle
exchange between them.

{\bf Multiple condensates.}
Consider a fluid of $N$ identical Bose particles.  Let $|0\rangle$ be
a  many-body state that is presumed to be both  (a) the ground state of
a many-body Hamiltonian ${\cal H}$ and  (b) a ``fragmented
condensate,'' {\em i.e.,} a state whose reduced one-body density matrix
\begin{equation}
\rho({\bf r}, {\bf r'}) \equiv \langle  \hat{\psi}({\bf r'})^\dagger
\hat{\psi}({\bf r})  \rangle
\end{equation}
has two or more macroscopic eigenvalues $N_i$, with corresponding 
eigenfunctions $\phi_i({\bf r})$. The $N_i$ have the physical
interpretation of the mean occupancy of the single-particle state
$\phi_i$ in the many-body state $|0\rangle$. In general the macroscopic
$N_i$ are not integers, and they need not add up to the total particle
number $N$. For simplicity, we will assume that  ony two such
macroscopic  eigenvalues exist, but our arguments are easily
generalized to any finite number.\cite{extra1}

We define the condensate annihilation operators 
\begin{equation}
b_i \equiv \int d{\bf r}\ \phi_i({\bf r}) \hat{\psi}({\bf r})
\end{equation}
where $\hat{\psi}({\bf r})$ is the particle annihilation operator
at position ${\bf r}$.  Then in the fragmented state,
\begin{equation}
[\hat{\rho}_0]_{ij}=
\langle 0 | b_j^\dagger b_i | 0 \rangle = N_i \delta_{ij} ,
\end{equation}
which follows from the assumption that $\phi_i$ are eigenstates of
$\hat{\rho}$.  For $i \neq j$ the density matrix element vanishes,
indicating the absence of coherence between the condensates.

{\bf A family of states.}
We now construct a family of $N$-particle states labeled by the integers
$q = 0, \pm 1, \pm 2, ...$ in which $q$ particles are transferred between
the two condensates.  For positive $q$, the transfer is from 1 to 2; for 
negative $q$, the transfer is from 2 to 1; the $q=0$ state is simply 
the original fragmented state $|0\rangle$.  These states can be formed by 
repeated applications of the operator
\begin{equation}
B^\dagger \equiv \frac{b_2^\dagger b_1}{\sqrt{N_2 N_1}}
\end{equation}
for positive $q$, and its Hermitian conjugate $B$ for negative $q$.  The 
factors in the denominator make the states
\begin{eqnarray}
| q \rangle &\equiv& (B^\dagger)^q | 0 \rangle \quad q \geq 0 \\
| q \rangle &\equiv& (B)^{|q|} | 0 \rangle \quad q < 0 
\end{eqnarray}
normalized for $q \ll N_i$, which will be the case of interest below.
The operator $\hat{Q}$ can then be defined as $\sum q|q\rangle\langle
q|$. Our strategy will be to construct a linear combination of the
states    $|q\rangle$ that is lower in energy than the putative ground
state $| 0 \rangle$.

{\bf An effective tight-binding model.}
We consider  a many-body Hamiltonian  ${\cal H}$ given by the sum of a
one-body part $\hat{h}_o$ -- kinetic energy plus a single-particle
potential $V({\bf r})$ -- and a two-body interaction $U({\bf r}-{\bf
r}')$, which for notational simplicity we will take to be a
$\delta$-function pseudopotential of strength $U$. (Our results are
easily extended to more general interactions.) Within the subspace of
states  $\{ | q \rangle \}$, ${\cal H}$ can be written
\begin{equation}
{\cal H} = -[ t_1 B^\dagger + t_1^* B ] -
 [t_2 (B^\dagger)^2 + t_2^* B^2] + \frac{1}{2}\kappa Q^2 ,
\label{tbm} 
\end{equation}
which has the form of a one-dimensional tight-binding model with 
nearest and next-nearest neighbor hopping matrix elements $t_1$ and
$t_2$, and a harmonic potential of spring constant $\kappa$.  

The generally complex hopping matrix elements are given by 
\begin{eqnarray}
t_1 &=& - \sqrt{N_1 N_2} \int d{\bf r}\ 
\phi_2^*({\bf r}) \hat{h}_{\rm eff} \phi_1({\bf r}) ,
\label{t1}
\\
t_2 &=& - \frac{U}{2} N_1 N_2 \int d{\bf r} 
\phi_2^*({\bf r}) \phi_2^*({\bf r}) 
\phi_1({\bf r})\phi_1({\bf r}) ,
\label{t2}
\end{eqnarray}
where $\hat{h}_{\rm eff} \equiv \hat{h}_o + 
U \sum_i N_i |\phi_i({\bf r})|^2$ 
is an effective one-body Hamiltonian that includes mean field
interactions.

If the condensates $\phi_i({\bf r})$ extend over the same
$d$-dimensional volume $R^d$, where $R$ is a characteristic length
scale, then the matrix elements of eq. (\ref{t1}\ref{t2}) scale as $t_1
\sim N \epsilon_1$ and  $t_2 \sim N \epsilon_2$, where the $\epsilon_i$
are functions of the mean density  $N/R^d$ with the units of energy per
particle.  Typically, they will be of the same order of magnitude as the
chemical potential.\cite{extra2}
We consider below the special case in which the
matrix elements $t_1$ and $t_2$ vanish for symmetry reasons.

The hopping terms of the Hamiltonian (\ref{tbm}) give rise to a 
tight-binding band with dispersion 
\begin{equation}
E(k) = -[t_1 e^{-ik}+t_1^* e^{ik}]-[t_2 e^{-i2k}+t_2^* e^{i2k}],
\label{band}
\end{equation}
which can be expanded about its minimum to yield
\begin{equation}
E(k) = -\frac{TN}{2} + \frac{T'N}{2}(k-\chi)^2 + ...
\end{equation}
where the real numbers $T$ and $T'$ depend on the $\epsilon_i$, and 
$\chi$ can be found by minimizing eq. (\ref{band}).  The effective mass
for this band is given by $\hbar^2/M_{\rm eff} = T'N$.

The ``spring constant'' $\kappa$ in eq. (\ref{tbm}) can be found quite generally
from the  assumption that the condensates are compressible, which is
expected on general grounds for a Bose fluid.\cite{extra3}
Then the energy of a
state with $q_i$ particles added to condensate $i$ (by repeated
application of $b_i^\dagger$) will be a smooth function that can be
Taylor expanded for small $q_i \ll N_i$, so that
\begin{equation} 
\langle q | {\cal H} | q \rangle = E_{0} + 
q( \mu_1 - \mu_2) 
+ \frac{\kappa}{2}q^2 + ...
\label{exp}
\end{equation}
where $\mu_i \equiv \partial E /\partial N_i$ is the chemical potential
of the $i$-th fragment and
\begin{equation}
\kappa = \frac{K}{N} \equiv [  \frac{\partial^2 E}{\partial N_1^2} - 
2\frac{\partial^2 E}{\partial N_1\partial N_2} + 
\frac{\partial^2 E}{\partial N_2^2} ]. 
\end{equation}
For condensates $\phi_i$ of volume $R^d$, $K$ is a number of order
$U(N/R^d)$ that is again comparable to the chemical potential, and hence
$T$ and $T'$.  Since by assumption $q=0$ is the ground state the
chemical potentials must be equal and the linear term in $q$ vanishes. 
For stability, $\kappa$ must be positive. 

The tight-binding model (\ref{tbm}) for two coupled condensates is
related by a  canonically transformation to the familiar pendulum
description of a Josephson  junction,\cite{LesHouches} which treats the
relative phase as a periodic ``coordinate'' and the number fluctuation
as its corresponding ``momentum.''  The present analysis instead takes
the number fluctuation as a discrete coordinate, with relative phase
corresponding to the conjugate (crystal) momentum. The two approaches
are completely equivalent, and provide complementary insights to the
behavior of coupled condensates.

{\bf A lower energy state.}
Within a continuum approximation, the tight-binding Hamiltonian
(\ref{tbm}) describes a particle of mass $M_{\rm eff}$ in a harmonic
potential of spring constant $\kappa$.  The frequency of this 
oscillator is given by $M_{\rm eff}\omega_{\rm eff}^2 = \kappa$, so 
that $\hbar\omega_{\rm eff} = \sqrt{T' K }$.  The ground state has the 
form
\begin{equation}
|G\rangle = \sum_{q} \frac{e^{-q^2/2\sigma^2}}{\pi^{1/4} \sigma^{1/2}} 
e^{i\chi q} | q \rangle,
\label{newstate}
\end{equation}
with a characteristic spread in $q$
\begin{equation}
\sigma = (\hbar/M_{\rm eff}\omega_{\rm eff})^{1/2} = (T'/K)^{1/4} 
N^{1/2} .
\end{equation}
Since $\sigma \sim N^{1/2}$, we are justified {\em a posteriori}
in both the continuum approximation and our decision to neglect terms 
of order $q^3$ (and higher) in $E(q)$.

The energy of $|G\rangle$ relative to $|0\rangle$ is
\begin{equation}
E_G - E_{0} = -\frac{TN}{2} + \frac{(T'K)^{1/2}}{2},
\end{equation}
where the first term on the right is the delocalization energy of the
bottom of the tight-binding band, and the second is the zero-point
energy of a harmonic oscillator with energy spacing $\hbar\omega_{\rm
eff}$.   Evidently the admixture of correlated condensate fluctuations
$|G\rangle$ is lower in energy than the fragmented state $|q=0\rangle$
if  $N > T'K/(T^2)$, where the right hand side is of order unity.\cite{extra4} 
Thus even for extremely weak off-diagonal couplings, a fragmented condensate
will not be the ground state for a macroscopic number of particles.  

{\bf Density matrix revisited.}
What is the nature of the true ground state $|G \rangle$?  It is easy to 
show that $|G \rangle$ is, in fact, a conventional Bose condensate whose
density matrix has a {\em unique} macroscopic eigenvalue.  Since
$q \ll N_i$, the diagonal matrix elements of $\hat{\rho}$ are unchanged
\begin{equation}
[\hat{\rho}_G]_{ii} = \langle G | b_i^\dagger b_i | G \rangle = N_i  .
\end{equation}
The off-diagonal matrix elements, however, are now macroscopic:
\begin{equation}
[\hat{\rho}_G]_{12} = \langle G | b_2^\dagger b_1 | G \rangle = 
g\ e^{-i\chi}\sqrt{N_1 N_2} 
\label{oddme}
\end{equation}
where
\begin{equation}
g = e^{-1/4\sigma^2} \approx 
1 - \frac{A}{N} ,
\label{g-def}
\end{equation}
where $A$ is a number of 
order unity.

Rediagonalizing the density matrix, we find a {\em single} 
macroscopically occupied eigenstate 
\begin{equation}
\phi_{\rm g}({\bf r}) = c_1 \phi_1({\bf r})
+ e^{i\chi} c_2 \phi_2({\bf r}),
\label{newcond}
\end{equation}
where
$c_i \equiv \sqrt{N_i/[N_1+N_2]}.$
The occupation of this state is $N_g = N_1 + N_2 - B$, where $B$
is proportional to $A$, and is also a number of order unity.
The orthogonal combination 
$\phi_{-}({\bf r}) = c_2 \phi_1({\bf r})
- e^{i\chi} c_1 \phi_2({\bf r})$
has eigenvalue $B$ of order unity.  We conclude that fragmented 
condensates are unstable towards the formation of conventional Bose 
condensate with a unique macroscopically occupied state.

{\bf Spontaneous symmetry breaking.}
As noted above, the matrix elements $t_i$ may vanish for symmetry
reasons. As a specific example, consider a rotationally invariant
system with two fragments $\phi_1$ and $\phi_2$ that have  angular
momentum projections $m_1$, $m_2$, respectively. Then the $t_i$ will be
zero.  This situation arises in the context of rotating Bose gases
trapped in an azimuthally symmetric  potential.\cite{Wilkin}  Under
these circumstances, we must ask if the fragmented condensate remains
stable in the presence of a small symmetry-breaking perturbation,
which introduces a nonzero $t_1$.  

We have seen, however, that even for $T \sim 1/N^2$ a fragmented
condensate will be unstable to the formation of a conventional
condensate (\ref{newcond}) that is a superposition of the fragments,
with a relative phase $\chi$ that is determined by the details of the
perturbing potential through eq. (\ref{t1}). The resulting state
is not an eigenstate of angular momentum, and therefore has an
asymmetric density and current distribution.\cite{longrbg}  The fact
that an order $1/N$ perturbation can reduce the symmetry of the ground
state indicates that fragmented condensates will spontaneously break
whatever symmetry (rotation, in the case of ref. \onlinecite{Wilkin})
permitted fragmentation in the first place.

This result can be understood in another way by considering the
susceptibility of a fragmented condensate to a perturbation which 
couples to the inter-fragment current density $\hat{J} \equiv$  $ i[
t_{1} B^\dagger - t_1^* B]$.  Since $\langle (B^\dagger)^2 \rangle = $
$\langle B^2 \rangle = 0$  in the $q=0$ state, while $B^\dagger B = B
B^\dagger = 1$ as an operator identity, the mean-square current
fluctuation $\langle J^2 \rangle \sim$ $|t_{1}|^2$.  A weak  one-body
potential that allows scattering between $\phi_1$ and $\phi_2$, however,
will typically introduce a $t_1 \sim N$. Fragmented condensates are thus
highly susceptible to perturbations that permit coherent particle
transfer between the fragments.

{\bf Gauge covariance.}
The phase factor $e^{i\chi}$ which enters into the superposition 
(\ref{newcond}) is not arbitrary, but is determined by the phase of the
off-diagonal density  matrix element (\ref{oddme}), which in turn is
given by the phase of the Hamiltonian matrix elements  $t_1$ and $t_2$
that scatter particles between the two condensates.  We can easily
confirm that this result is  properly gauge covariant.\cite{Anderson}

The single-particle states $\phi_i$ are eigenstates of the (fragmented)
one-body density matrix $\hat{\rho}_0$, and are only defined to within
arbitrary phase factors.  All physical observables must therefore be
unchanged if each $\phi_i$ is  multiplied by $e^{i\alpha_i}$.  It is
easy to follow these phases through our calculations.  The hopping 
matrix elements transform as $t_1 \rightarrow t_1
e^{i(\alpha_1-\alpha_2)}$ and $t_2 \rightarrow t_2
e^{2i(\alpha_1-\alpha_2)}$, so that $\chi
\rightarrow \chi + \alpha_1 - \alpha_2$.   At the end of the day, the
unique condensate (\ref{newcond}) becomes
\begin{eqnarray}
\phi_{\rm g}({\bf r}) &\rightarrow& c_1 [e^{i\alpha_1} \phi_1({\bf r})]
+ e^{i(\chi + \alpha_1 - \alpha_2)} [ c_2 e^{i\alpha_2} 
\phi_2({\bf r})] \nonumber \\
&=& e^{i\alpha_1} \phi_{\rm g}({\bf r}) .
\end{eqnarray}
The arbitrary phases $\alpha_i$ are only reflected in the overall phase
of $\phi_{\rm g}$  (which itself is only defined up to an overall
phase).  The {\em relative} phase of $\phi_{\rm g}({\bf r})$ between any
two points ${{\bf r}_1}$  and ${{\bf r}_2}$ are invariants.  The 
superfluid velocity, which is proportional to the gradient of the phase,
is also invariant.  Thus the phase of the off-diagonal one-body density
matrix element (proportional to $\langle B \rangle$)  is a
gauge-covariant representation of the relative phase between two
condensates.  

{\bf Relative phase of previously isolated condensates.}
We can use a similar approach to analyse Anderson's   {\em
gedankenexperiment}\cite{Lesson} in which two independently created
condensates $\phi_L$ and $\phi_R$ are placed in contact.  Before contact
is established, the many-body state of the combined system is a simple
product of states -- {\em i.e.,} a ``fragmented condensate'' -- that we
can denote $|q = 0\rangle$.  Since the off-diagonal element $\langle q=0
| b_L^\dagger b_R | q=0 \rangle$ of the density matrix is zero, the
relative phase between the two condensates is the phase of the complex
number zero, which is ill-defined.  

For {\em gedanken}purposes, let us assume that the two independent
condensates are identical, with equal chemical potentials. (Our
discussion is easily generalized.)   As shown in the previous section,
we may without loss of generality choose both $\phi_L$ and $\phi_R$ to
be real.  The time evolution following the initiation of contact is
governed by a Hamiltonian of the form (\ref{tbm}), plus
damping,\cite{LesHouches} with $t_2 << t_1$. By time-reversal symmetry,
the $t_i$ are real, and $\chi = 0$ or $\pi$; for definiteness we assume
conventional tunneling contact, with $\chi = 0$. In contrast to the case
of spatially overlapping fragments, however, the Josephson coupling
$t_{\rm eff}$ between two reservoirs in contact will be proportional to
their contact area, and exponentially small in the energy barrier between
them.\cite{extra2}

Except for the discreteness of $q$, this is a familiar problem.  At long
times, the system ends up in a Gaussian ground state, which we have seen
is simply a conventional condensate with uniform phase.  Relative to
this ``vacuum,'' the initial state $q=0$ is ``squeezed,'' with a width
$\delta q \sim 1$ in position and $\delta p \sim \pi$ in the conjugate
momentum. The establishment of phase coherence between two previously
isolated condensates is therefore equivalent to the decay of a
``squeezed vacuum'' in quantum optics.\cite{Walls}  For our purposes,
the important feature is that due to the $\pm q$ symmetry of the initial
state and the Hamiltonian,  $\langle B^\dagger \rangle = \langle B
\rangle$ for all time.  The off-diagonal density matrix element is then
always real, so as soon as the relative phase becomes defined, it is
zero.

The experiment of Andrews {\em et al.}\cite{Andrews} in which two
nominally identical, initially isolated condensates are allowed to
expand and overlap can also be cast in this form. If the condensates
are initially converging symmetrically, then $\phi_R(z) =
\phi_L(-z)^*$, and we again find that the transfer matrix element  that
scatters particle between the condensates is real.  This result is
consistent with Naraschewski {\em et al.}'s analysis\cite{Nara} of the
interference fringes seen in the MIT  experiment, which requires $\chi =
0$.

{\bf Random vs. reproducible phases.} 
Javanainen and Yoo \cite{JY} and Castin and Dalibard \cite{CD} have
discussed  protocols in which a series of measurements is performed on
the states of  individual particles removed from two isolated
condensates. Their elegant analyses show that the interference patterns
that emerge are consistent with a well-defined phase difference between
the two condensates.  This phase difference, however, is arbitrary, and
varies if the entire series of measurements are repeated. How can our
calculation be reconciled with these results?

In the schemes of refs. \onlinecite{JY} and \onlinecite{CD}, each
measurement of the series removes a particle from the two-condensate
system in a defined superposition of the original condensates $\phi_i$.
The remaining particles are therefore left in a specific entangled 
many-body state that depends on the results of the measurements. In
other words, a series of observations of the removed particles generates
a specific perturbation $\delta {\cal H}(t)$ acting on the remaining
particles. This perturbation does not conserve particle number, and
in effect applies a time-dependent, gauge-symmetry-breaking 
``$\eta$-field''\cite{Hohenberg} to the condensates. 

As we have seen, the susceptibility of the state $q=0$ to such a
perturbation is divergent with $N$.  Thus it only takes a small
perturbation ({\em i.e.}, a few measurements) to sculpt the state of the
remaining particles into a single condensate, whose relative phase
depends deterministically on the $\delta {\cal H}(t)$, and hence on the
specific sequence of observations performed on the removed particles. We
have shown above that when the two condensates are placed in
particle-exchanging contact their coupling {\em also} introduces a
perturbation, which reproducibly enforces a relative phase of zero.  An
interesting question is the manner in which this effect competes with
the measurement scenarios of refs. \onlinecite{JY} and \onlinecite{CD}.

\noindent{\bf Acknowledgements.} I thank N. Wilkin, R. Smith, J. Gunn, D. Butts,
M. Mitchell, J. Ho,  J.C. Davis, D. Weiss, J. Garrison, S. Kivelson, and R.
Chiao for many interesting conversations.


\begin{references}

\bibitem{Penrose}
O. Penrose, 
{\em Phil. Mag.} {\bf 42}, 1373-1377 (1951).

\bibitem{Beliaev}
S.T. Beliaev, {\em J. Exp. Theor. Phys. (USSR)} {\bf 34}, 417-432
(1958).

\bibitem{Yang}
C.N. Yang, {\em Rev. Mod. Phys.} {\bf 34}, 694 (1962).


\bibitem{Anderson}
P.W. Anderson, in {\em Lectures on the Many Body Problem},
ed. E.R. Caianello (Academic Press, New York 1964) vol. 2, 113.

\bibitem{Anderson2}
P.W. Anderson, {\em Some Recent Definitions in the Basic Sciences}, vol
2, ed.  A. Gelbert,  (Belfer Graduate School of Science, Yeshiva
University, NY 1965-6) 21-40.

\bibitem{Nozieres-StJames}
P. Nozieres and D. Saint James, J. Phys. (Paris) 43, 1133 (1982).

\bibitem{Nozieres}
P. Nozieres,  
in {\em Bose-Einstein Condensation},  ed. A. Griffin, D.W. Snoke, S.
Stringari (Cambridge University Press, 1995), 15-30. 

\bibitem{Wilkin}
N. K. Wilkin, J. M. F. Gunn, and R. A. Smith.
{\em Phys. Rev. Lett.} {\bf 80} 2265-2268 (1998).

\bibitem{Pethick}
O. Elgaroy and C.J. Pethick, cond-mat 9805144.

\bibitem{Lesson}
P. W. Anderson, in {\em The Lesson of Quantum Theory}, 
J. D. Boer, E. Dal, O. Ulfbeck, Eds. (Elsevier, Amsterdam, 1986), pp. 23-33.

\bibitem{LeggettSols}
A. J. Leggett and F. Sols, {\em Found. Physics} {\bf 21}, 353 (1991). 

\bibitem{Leggett}
A.J. Leggett, in {\em Bose-Einstein Condensation}.  ed. A. Griffin, D.W. Snoke, S.
Stringari. (Cambridge University Press, 1995) pp. 452-462.

\bibitem{JY}
J. Javanainen and S. M. Yoo, {\em Phys. Rev. Lett.} 76, 161 (1996). 

\bibitem{CD}
Y. Castin and J. Dalibard, {\em Phys. Rev.} {\bf A} {\bf 55}, 4330 (1997). 

\bibitem{Wallisetal}
H. Wallis, A. Rohrl, M. Naraschewski, A. Schenzle, and H.J. Miesner,
{\em J. Mod. Optics,} {\bf 44}, 1751 (1997).

\bibitem{Villain}
P. Villain, M. Lewenstein, R. Dum, {\em et al.}
{\em J. Mod. Optics,} {\bf 44}, 1775 (1997).


\bibitem{longrbg}
D.A. Butts and D.S. Rokhsar, ``Predicted signatures of rotating 
Bose-Einstein condensates,'' to appear, Nature (1998).

\bibitem{akin}
The next terms in the expansion (\ref{exp}) are of the form $q^3
(\partial^3 E /\partial N_i\partial N_j\partial N_k)$.  For a large
system, the  third derivative should vary as  $N^{-2}$ times an
intensive quantity.  Since we will be interested in $q < (N)^{1/2}$,
such terms will vary as $N^{-1/2}$ and can be neglected
for large systems.  The same argument holds for higher orders.

\bibitem{LesHouches}
A.J. Leggett, in {\em Chance and Matter} (North-Holland, 1987), p 395.

\bibitem{Walls}
D.F. Walls and G.J. Milburn, {\em Quantum Optics.} (Springer-Verlag,
1994).

\bibitem{Andrews}
M.R. Andrews, C.G. Townsend, H.-J. Miesner, {\em et al.}
{\em Science} {\bf 275}, 637 (1997). 

\bibitem{Nara}
M. Naraschewski, H. Wallis, A. Schenzle, J. I. Cirac, ibid. 54, 2185
(1996).  J. I. Cirac, C. W. Gardiner, M. Naraschewski, P. Zoller,
ibid., p. R3714. 

\bibitem{Hohenberg}{
P.C. Hohenberg and P.C. Martin, {\em Ann. Phys.} {\bf 34}, 291 (1965).
}

\bibitem{extra1}
In a straightforward extension of the analysis given here, the case of
$N_{\rm con}$ coexisting condensates can be mapped onto an 
$(N_{\rm con}-1)$-dimensional harmonic oscillator, and the unique
ground state can be determined.  This many-body ground state is a conventional
Bose condensate with a unique condensate wavefunction.  The crucial requirement
is that the occupation numbers of the condensates must be macroscopic, hence
the restriction to a finite number of condensates.  It is an interesting
question to ask whether the number of condensates $N_{\rm con}$ can scale
in some nontrivial way with total particle number $N$ to preserve
fragmentation.

\bibitem{extra2}
We emphasize that the behavior of multiple spatially {\em overlapping} 
condensates (as proposed by Ref. \onlinecite{Wilkin}, and considered in the first 
part of the present work) can be different from that of two spatially  {\em
separated} condensates (whose behavior when brought into contact is discussed
in the second part of this paper).  For  separated condensates the hopping
matrix elements $t_i$ will be exponentially  small in the height of the
intervening energy barrier.  Then $TN$, while formally macroscopic ({\em i.e.},
proportional to $N$), can also be negligibly small.  For overlapping
condensates this behavior is not expected. 

\bibitem{extra3}
The assumption that the condensate fragments are compressible fluids
excludes the case of Bose insulators in particle-exchanging contact, for 
which the energy $\langle q |{\cal H} | q \rangle$ would depend on $|q|$. 

\bibitem{extra4}
We emphasize again that for the case of two condensates separated by a 
barrier, $T$ and $T'$ will be exponentially small in the barrier height. Thus
according to the condition derived here, a ``fragmented'' state could then be
stable even for ``macroscopic'' $N$.  This is simply the statement that two
sufficiently far separated reservoirs of bosons can condense independently.
This behavior is distinct from the fragmentation of spatially overlapping 
condensates as discussed in ref. \onlinecite{Wilkin}.



\end{references}
\end{document}